\documentclass{bmcart}

\RequirePackage{hyperref}
\usepackage[utf8]{inputenc} 
\usepackage{url}
\usepackage{graphicx}
\usepackage{color}

\startlocaldefs
\endlocaldefs

\begin{document}

\begin{frontmatter}

\dochead{Research}


\title{Modeling the rise in Internet-based petitions}


\author[
   addressref={aff1},                   
  corref={aff2},                       
]{\inits{T}\fnm{Taha} \snm{Yasseri}}
\author[
   addressref={aff1},
]{\inits{S}\fnm{Scott A.} \snm{Hale}}
\author[
   addressref={aff1},
]{\inits{HZ}\fnm{Helen Z.} \snm{Margetts}}


\address[id=aff1]{
  \orgname{Oxford Internet Institute, University of Oxford}, 
  \street{1 St Giles'},                     %
  \postcode{OX1 3JS}                                
  \city{Oxford},                              
  \cny{UK}                                    
}


\begin{artnotes}
\note{Oxford Internet Institute, University of Oxford, OX1 3JS Oxford, UK}     
\end{artnotes}



\begin{abstractbox}

\begin{abstract} 
Contemporary collective action, much of which involves social media and other Internet-based platforms, 
leaves a digital imprint which may be harvested to better understand  the dynamics of mobilization. 
Petition signing is an example of collective action which has gained in popularity with rising use of social 
media and provides such data for the whole population of petition signatories for a given platform. This paper 
tracks the growth curves of all 20,000 petitions to the UK government over 18 months, analyzing the rate of growth 
and outreach mechanism. Previous research has suggested the importance of the first day to the ultimate success of a 
petition, but has not examined early growth within that day, made possible here through hourly resolution in the data. 
The analysis shows that the vast majority of petitions do not achieve any measure of success; over 99 percent fail to 
get the 10,000 signatures required for an official response and only 0.1 percent attain the 100,000 required for a 
parliamentary debate. We analyze the data through a multiplicative process model framework to explain the heterogeneous 
growth of signatures at the population level. We define and measure an average outreach factor for petitions and show that 
it decays very fast (reducing to 0.1\% after 10 hours). After 24 hours, a petition's fate is virtually set. The findings 
seem to challenge conventional analyses of collective action from economics and political science, where the production 
function has been assumed to follow an S-shaped curve.
\end{abstract}



\end{abstractbox}
%

\end{frontmatter}



\section*{Introduction}
Increasingly collective action takes place in whole or at least in part online \cite{harlow2012}, leaving transactional data that allow for new forms of analysis. Studying the dynamics of protest recruitment through social media \cite{gonzalez2011,borge2011,ghonim2012revolution}, modeling emergence and resolution of conflict in online 
mass collaboration projects \cite{yasseri2012,iniguez2014}, characterizing online partisans \cite{conover2012}, and 
quantifying ``collective emotions'' \cite{chmiel2014} are some examples of the research based on these new type of data, which have been called ``big data'' in the literature. 

Petition-signing provides an example of a popular, low-cost act of political 
participation that is increasingly carried out digitally and shared via social media \cite{hale2014}. 
In this paper, we analyze growth patterns in petitions submitted to the UK Government on the central government portal. This electronic petition 
platform was developed by the UK Cabinet Office for the Coalition Government in 2010 and launched in August 2011 at \url{http://epetitions.direct.gov.uk}.

\begin{figure}[h!]\includegraphics[width = .7 \linewidth]{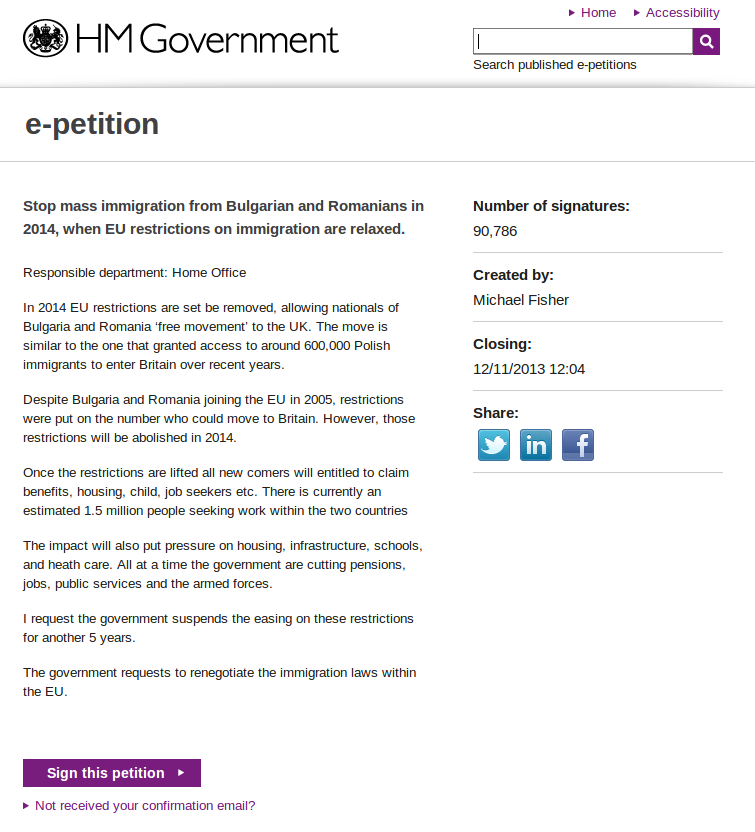}
  \caption{\csentence{A snapshot of a petition on the \url{http://epetitions.direct.gov.uk} site.}}\label{fig:screenshot}
      \end{figure}

The petition platform replaced an earlier government platform on the No.~10 Downing Street website, which was the first online petition
platform operated by the UK government.
The previous platform ran from November 2006 until March 2011, during which time the site received more than 12 million signatures
associated with over 5 million unique email addresses \cite{wright2012}.
Both the No.~10 site and the newer Cabinet Office site have allowed anyone to
view petitions and any user with a valid email address and UK postcode to create a new petition or to sign an existing petition. There are important
differences between the sites, however. Whereas the first site showed the names of the 500 most recent signatories to a petition, the 
new site shows only the name of the petition creator (see Figure~\ref{fig:screenshot}). The sites also have provided alternative 
measures for the ``success'' of a petition. For the earlier, No.~10 site, the government guaranteed an official response to all 
petitions receiving at least 500 signatures, while for the new site the coalition government promises an official response at 10,000 
signatures and that any petition attracting more than 100,000 signatures will qualify for a parliamentary debate  on the issue raised.

In this paper we focus on the second petition website. In previous work \cite{hale2013-websci} we analyzed data from the first website, finding that
the number of signatures a petition received on its first day was pivotal to its ultimate success. The low threshold for success 
(500 signatures) and the coarse daily resolution of data on growth for the earlier site did not allow for an in-depth examination of the 
critical early moments of petitions. In this paper, we undertake a new in-depth investigation of the early growth of petitions aided by the higher 
threshold for success on the new platform (100,000 vs 500 signatures) and a finer-grained capture of petition growth (hourly vs.~daily resolution). 
Analysis of this second platform also allows us to compare its dynamics with those of the earlier platform, which we do throughout our analysis where possible, to find properties of petition growth that transcend any specific web platform

\section*{Background}
Signing petitions has long been among the more popular political activities, leading the field for
participatory acts outside of voting \cite{parry1992}.
In addition to the potential to bring about policy change, petition signing has had other social benefits ascribed to it such as 
reinforcing civic mindedness \cite{whyte2005}. Online petitioning is one of a growing portfolio of Internet-based democratic innovations 
\cite{smith2009}, and the widespread use of electronic petition platforms by both governments and NGOs (e.g., Avaaz and 38 Degrees) has
received accolades for their democratic contribution \cite{escher2011, chadwick2008}.

The UK petition platforms have received little attention in recent political science research,
with the exception of qualitative work by Wright \cite{wright2012,Hansard}. The German e-petition platforms have been
analyzed more extensively \cite{lindner2011,jungherr2010}.
Jungherr and J\"{u}rgens \cite{jungherr2010} examined the distribution of signatures to the top petitions and found that while some successful petitions had an early number of signatures others
received the bulk of their signatures later. Schmidt and Johnsen combined qualitative and quantitative methods to provide a typology for the petitions based on their dynamics \cite{schmidt2014}.

Online petitions are examples of mobilizations with strong online imprints, which will include the entire transaction 
history for both successful and unsuccessful mobilizations. The data that can be harvested from the signing of electronic petitions 
represent a transactional audit trail of what people actually did 
(as opposed to what people think they did) and an entire population (without the need to take a representative sample). Data like 
this represents a shift for social science research into political behavior, which has traditionally rested on survey data, or, 
for elections, voting data.
These data make it possible to look at the different patterns of growth in the 20,000 mobilization 
curves that we have and identify the distinctive characteristics of those mobilizations that succeed and those that fail with our digital 
hindsight. Such studies, using data that has rarely been available to
political science researchers before the current decade, may tell us something about the nature of collective action itself in a digital 
world. Of the research noted above, Jungherr and J\"{u}rgens used a smaller dataset to illustrate the viability of a big data approach
(or computational social science) \cite{jungherr2010}, but other studies used surveys \cite{lindner2011} or more qualitative approaches \cite{wright2012}.

\section*{Data}
The UK Government's petition website was accessed hourly from 5 August 2011 to 22 February 2013 with an automated script. Every hour, the number of 
total signatures to date for each active petition was recorded. In addition, whenever a new petition appeared, the title of the petition, the name of the petitioner, the text of the petition, 
the launch date of the petition, and the government department at which the petition was directed were recorded. Overall, 
19,789 unique petitions were tracked, representing all active petitions available publicly at any point during the study.

\section*{Results}
\subsection*{Overall statistics}
\begin{figure}[h!]\includegraphics[width = .5 \linewidth]{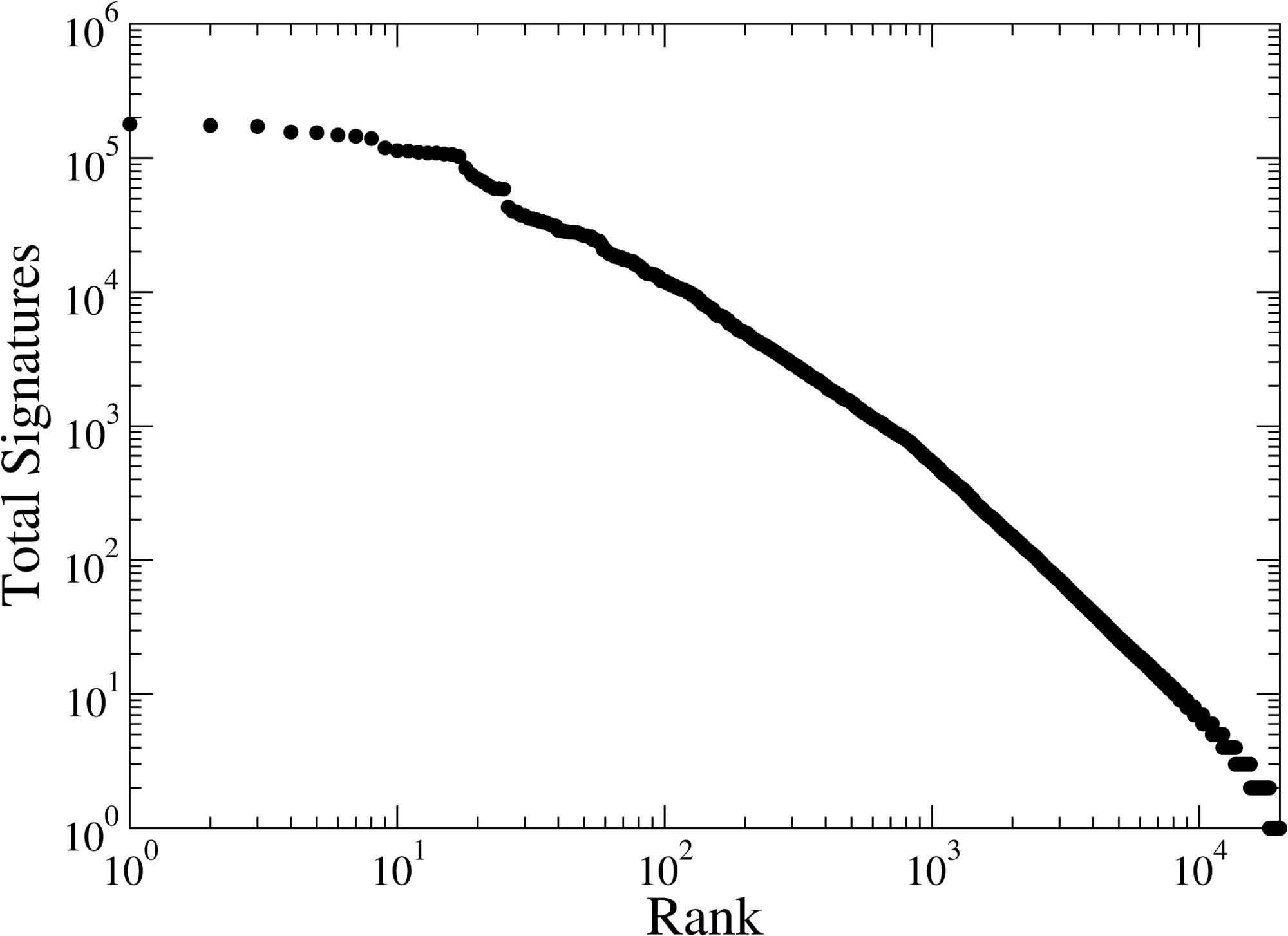}
  \caption{\csentence{Total number of signatures per petitions plotted against the rank order of the petition based on its total number of signatures.}}\label{fig:zipf}
      \end{figure}

A total of 7,303,019 signatures were collected for the 19,789 petitions. Figure~\ref{fig:zipf} shows that the distribution of these 
signatures is highly skewed,
by plotting the total number of signatures for each petition against the rank order of the 
petition by total number of signatures. It is clear that a small number of petitions have been signed many times each, while a large number of 
petitions have only been signed a few times each (indeed, half the petitions received only one signature).

Only 5 percent of petitions obtained 500 signatures in total, which is similar to the percentage achieving 500 signatures on the previous, 
No.~10 petition site \cite{hale2013-websci}. Beyond this,  4 percent of the petitions received 1,000 signatures. Only 0.7 percent attained 
the 10,000 signatures required for receiving an official response, and 0.1 percent attained the 100,000 signatures 
required for a parliamentary debate.

Despite the much larger threshold for success compared to the previous No.~10 platform (10,000 vs 500 signatures for an official response and
the additional measure of success of 100,000 signatures for a parliamentary debate), a similar pattern in growth 
emerges suggesting that the first day was crucial to achieving any kind of success. Any petition receiving 100,000 
signatures after three months, needed to have obtained 3,000 signatures within the first 10 hours on average. The external measure of 100,000 
signatures as success is also clear in Figure~\ref{fig:zipf}: petitions rarely grow further once passing the 100,000 signature mark.

\subsection*{Outreach and growth}
\begin{figure}[h!]\includegraphics[width = .9 \linewidth]{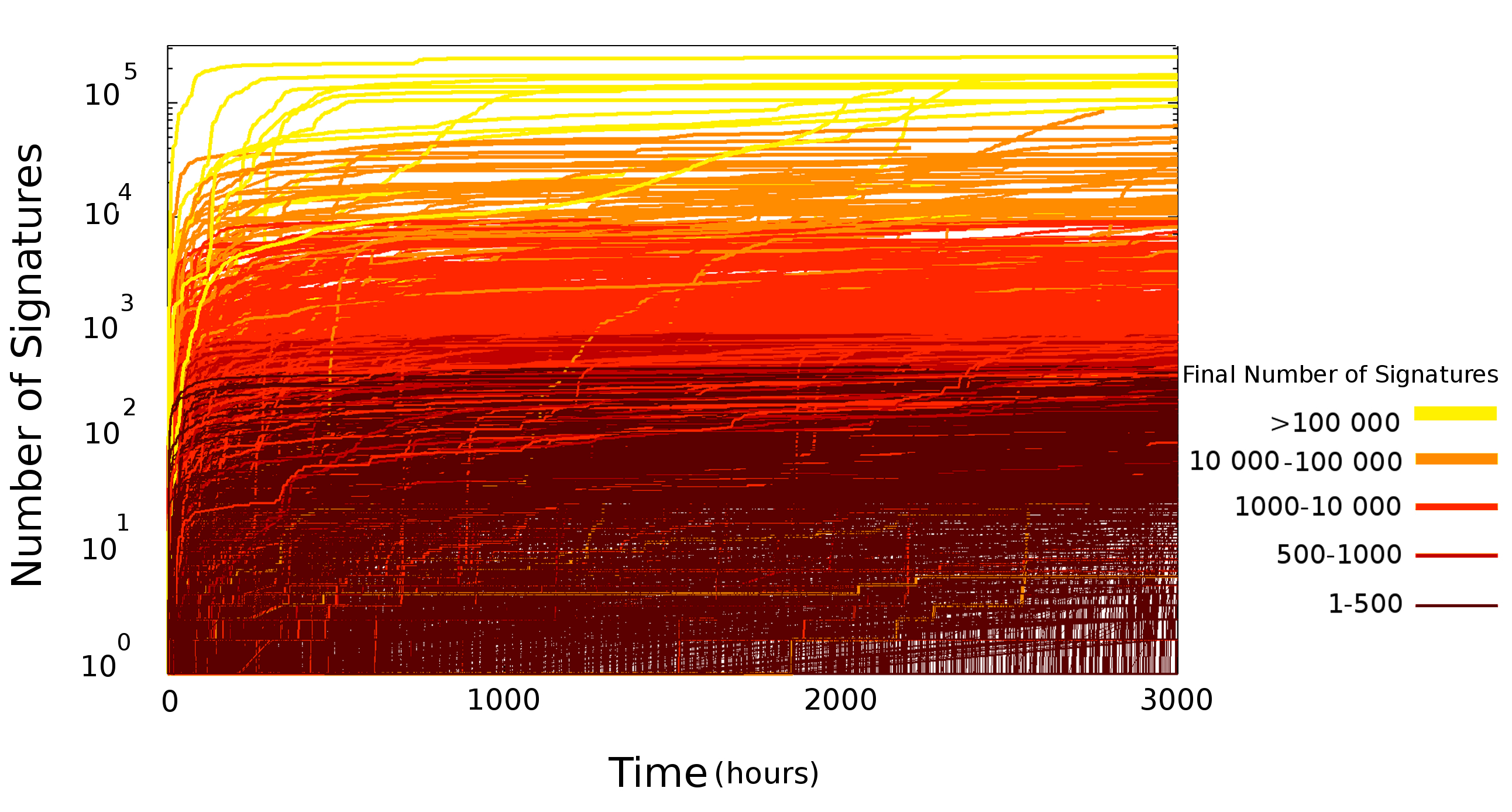}
  \caption{\csentence{Growth of the number of signatures for all petitions. Lines are shaded according to the 
  total number of signatures on the petition at the end of the collection period.}}\label{fig:growth}
      \end{figure}

Figure~\ref{fig:growth} shows the number of signatures over time (hours passed from each petition's launch time). 
The lines are shaded according to the final number of signatures each petition collected. From this figure it can be easily 
observed that even those petitions with large number of signatures, collected them very shortly after the 
launch and generally after a few days the growth slows significantly for all petitions (note the logarithmic scale of the y-axis).

At first glance these findings seem to contradict the normal assumptions of economists and sociologists, who have assumed the production function 
for collective action to follow an S-shaped curve, with the shape dependent upon the distribution of thresholds in the population \cite{granovetter1978,schelling1978,centola2013,marwell1993}. 
Rather than a slow accumulation of supporters building up to critical mass, at which point support ``tips over'' into success, petitions that have been successful in receiving large number of signatures demonstrate 
very rapid early growth, which decelerates overtime. The exponential growth of the number of signatures within the first few hours would suggest even a finer zoom into the data would not capture
any build-up period.
      
We attempt to capture the characteristic of early rapid growth and decay that the data reveal, with a model of ``collective attention''
decay, drawing on Wu and Huberman \cite{wu2007}. In their model, they calculate a ``novelty'' parameter relating to the novelty of news 
items on Digg (\url{http://digg.com}), a news sharing platform, that decays over time. In a more general framework, the decay in attention could occur for different reasons, for 
example reaching the system size limit, or lack of viral spread. 

In the model, $N$ agents at the time $t$, bring $N\mu$ new agents in the 
next step on average, $\mu$ being a multiplication factor. In our case, this would mean that every signature on petition $i$ brings $\mu_i$ 
new signatures in the next hour, leading to an exponential growth of rate $\mu_i$ in the number of signatures. This model fits the data we 
observe empirically quite well for the short period of time directly after a petition's launch (see Figure~\ref{fig:growth}). Very soon, however, the rate decays
and new signatures come at a much lower rate.

As in the model of Wu and Huberman \cite{wu2007}, we introduce a decay factor to capture this decrease. Specifically, we let the 
multiplication factor decay by introducing a second factor $r(t)$, which decays in a way that is intrinsic to the medium: each signature
at time $t$, on average brings $\mu_i r(t)$ new signatures in the next hour. This ``outreach parameter'' can change overtime and dampen the fast initial growth, correcting for the early saturation observed in the empirical data. The growth equation then reads:

\begin{eqnarray}
N_i(t+1) = N_i(t) (1+\mu_i r(t)) 
\end{eqnarray}

The number of signatures at time $t$ then can be written as:
\begin{eqnarray}
N_i(t) = N_i(0) (1+\mu_i r(0)) (1+\mu_i r(1)) \dots (1+\mu_i r(t-1))
\label{eq:N_t}
\end{eqnarray}
In the limit of small time increments, Equation~\ref{eq:N_t} converts to:
\begin{eqnarray}
N_i(t) = N_i(0) \mathrm{e}^{(\mu_i \sum_{t'=0}^{t'=t}(r(t'))}
\label{eq_exp}
\end{eqnarray}
We can assume that the number of signatures at the beginning is one (the initiator of the petition), and therefore averaging of the logarithm of both sides of Equation~\ref{eq_exp} leads to:
\begin{eqnarray}
\mathrm{E}[\mathrm{log}(N_i(t))] = \mathrm{E}[\mu_i] \sum_{t'=0}^{t'=t}(r(t')),
\end{eqnarray}
where $E[.]$ indicates the average over the whole sample.

In this framework, each petition has its own fitness and therefore an individual growth rate of $\mu_i$, whereas $r$ 
characterizes the overall outreach power of the platform as a whole. The outreach of the platform is assumed to be independent 
of the petition fitness and popularity. 
The disentanglement between these two factors enables us to calculate the outreach 
factor of the system by considering the whole sample of petitions and averaging over the logarithm of the number of signatures 
in hourly bins, starting from the time a petition is launched and then calculated in hourly increment at time $t$ and normalized 
by the logarithm of the number of signatures up to time $t$ as follows:
\begin{eqnarray}
r(t) = \frac{\mathrm{E}[\mathrm{log}(N_i(t))] - \mathrm{E}[\mathrm{log}(N_i(t-1))]}{\mathrm{E}[\mathrm{log}(N_i(t)]}.
\label{eq_r}
\end{eqnarray}

\begin{figure}[h!]\includegraphics[width = .8 \linewidth]{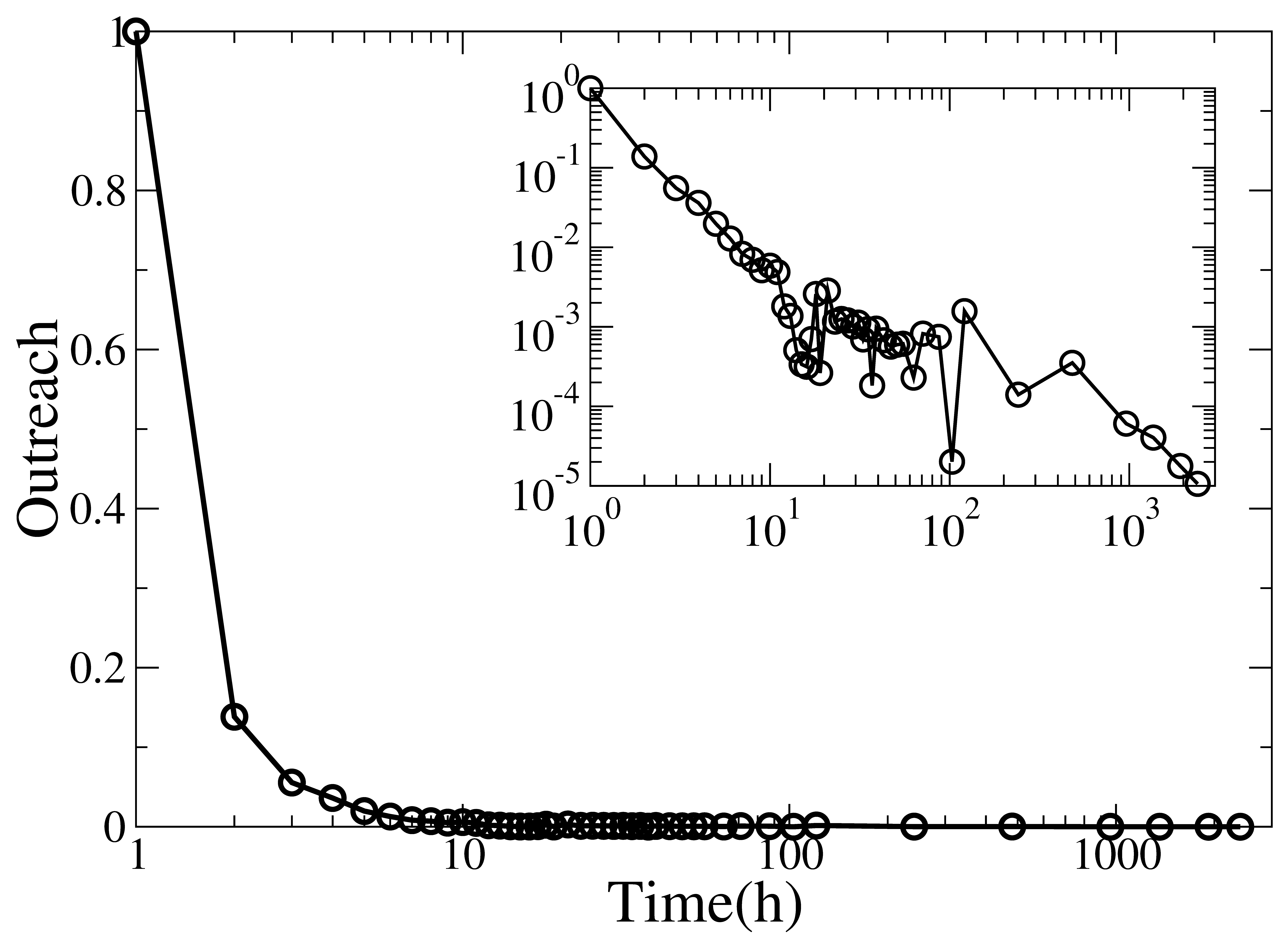}
  \caption{\csentence{The outreach factor for the petition site as a whole calculated according 
  to Equation~\ref{eq_r}. The inset shows the same quantity on a log-log scale.}}\label{fig:outreach}
      \end{figure}

We have calculated the outreach factor as a function of time according to Equation~\ref{eq_r} and illustrated it in Figure~\ref{fig:outreach}. 
The outreach factor decays very fast and, after a time span of 10 hours, reduces to 0.1\%.

\begin{figure}[h!]\includegraphics[width = .6 \linewidth]{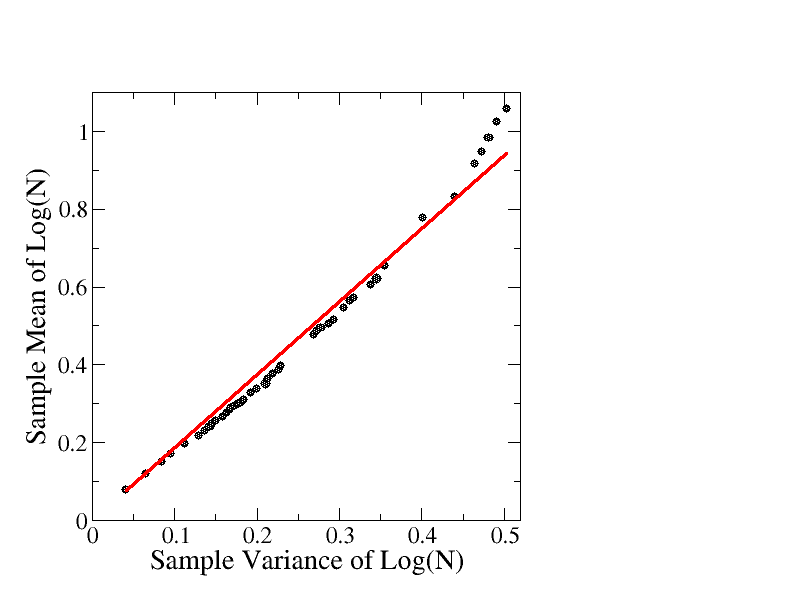}
  \caption{\csentence{The sample average of $\mathrm{log}(N)$ against the variance of the same quantity 
  to validate the multiplicative model according to Equation \ref{eq:mean_var}. The solid line is a linear fit of the form of $y=ax$ with $a = 1.872 \pm 0.03$.}}\label{fig:sample_avg}
      \end{figure}

The model holds, however, only when the growth rates of different petitions come from a localized  distribution with finite 
average and variance. To check this condition, we calculate the ratio between the sample average and variance 
of $\mathrm{log}(N(t))$ for different $t$ and check the following linear relation holds:
\begin{eqnarray}
\frac{\mathrm{E}[\mathrm{log}(N(t))]}{\mathrm{Var}[\mathrm{log}(N(t))]} = \frac{\mathrm{E}[\mu_i] \sum_{t'=0}^{t'=t}(r(t'))}{\mathrm{Var}[\mu_i] \sum_{t'=0}^{t'=t}(r(t'))} = \frac{\mu}{\sigma^2},
\label{eq:mean_var}
\end{eqnarray}
where $\mu$ and $\sigma$ are the sample average and the standard deviation of the individual growth rates. 
If the multiplicative model and the framework are valid, the ratio between the sample mean and the variance of $\mathrm{log}(N)$ should remain constant over time. 
Figure~\ref{fig:sample_avg} plots these two values and demonstrates the ratio does indeed remain constant. The root mean square of residuals from a diagonal line is 0.03.

\section*{Discussion and Conclusions}
This paper analyzes growth patterns in petitions created on the UK government petitions platform for 18 months from August 2011 to February 2013. 
We find that most petitions started on the platform failed to achieve any real traction, while the minority of petitions that did
amass a large number signatures did so quickly. The distribution of the number of signatures per petition is highly skewed: a few 
petitions captured a large number of signatures, while most petitions received very few signatures. By applying a simple multiplicative growth 
model, we have illustrated that the intrinsic time scale of the platform is very short and the growth of signatures on petitions 
exhibits rapid dynamics.

These findings have immediate application to petition platform designers as well as to petitioners themselves. Although the UK 
site defaults to having petitions active for one year, our analysis indicates that most signatures are added shortly after a
petition is launched. Shorter deadlines such as the three\slash{}six week deadlines of German petitions \cite{jungherr2010} or the one month 
deadline of US petitions on the WeThePeople site, therefore, might produce similar outcomes without the clutter of old petitions on the sites.
We are examining this design option through a comparison with the US and German petitions platforms, 
and advising policy-makers regarding the optimal design\footnote{\url{http://www.parliament.uk/business/committees/committees-a-z/commons-select/procedure-committee/inquiries/parliament-2010/e-petitions}}. The analysis also highlights the
importance for petitioners of gaining early traction. Experimental research shows that the willingness of individuals to sign a 
petition varies with the social information provided on how many other individuals have signed the petition already
\cite{margetts2011-epsr, margetts2013-ps}. The early growth of petitions reflects a similar feedback loop as the petitions with the most 
signatures get further signatures. The outreach factor, however, decays very quickly indicating that the window of opportunity for 
success is very small on the platform.

The outreach factor is fit to all the data on the platform and reflects the collective decay in attention to the platform as a 
 whole. In comparison to the work by Wu and Huberman \cite{wu2007} on the Digg news sharing website, where the outreach 
factor (there called novelty) decayed ``faster than power law,'' with a half-life of approximately one hour, the petition 
outreach factor decays very close to a power law. 

Thus, attention to petitions lasts somewhat longer than attention to the news 
links analyzed by Wu and Huberman \cite{wu2007}. It will be useful in future research to compare the outreach of various 
platforms, political and non-political, to understand the variations in the dynamics of different platforms. The growth patterns 
revealed by this analysis challenge our current understanding of collective action in online environments. 

The explanation for the different patterns observed here may derive from the means by which petition signatories come to find out about petitions.
Analytics data for the petition platform show the importance of social networking sites in disseminating 
petitions: about 50\% of users arrive at the petitions platform via either Facebook or Twitter, which also have a very high intrinsic pace \cite{bright2014}. 

Intuitively, we would expect dissemination via social networking sites to encourage early bursts, because if a user does not receive the information about 
a petition in their Facebook feed or Twitter stream when it is first sent and appears there, they will not do so unless they seek it out, which evidence from 
previous research suggests most users do not \cite{Lin2013,gleeson2013}.

Other investigations into the role of social 
networking sites in the growth of mobilizations have identified an S-shaped curve, with critical mass reached only after ``mid'' participants have responded 
to evidence of the contribution of early participants \cite{gonzalez2011}. 
Such accounts follow conventional frameworks of collective action in assuming that people have 
heterogeneous thresholds for joining a mobilization and that until participation reaches the 
threshold of the majority of potential participants, the joining rate will be slow \cite{Granovetter, 
schelling1978}.
In contrast, we find no slow early growth in those mobilizations that were 
successful, but rather a very rapid growth such that even with the hourly resolution of data collection, we do not observe any initial slow growth phase.

In the context of petitions, it seems that becoming aware of a petition is more important to eventual 
success (analogous to sharing news, as in Wu and Huberman~\cite{wu2007}) than influence from the originator of 
the petition or the identity of those who have shared it on social networking platforms (as in, for example, \cite{gonzalez2011}). 
Signing a petition is a very low cost activity (especially when the name of the signer is not publicly displayed). As such there are many individuals with low enough
thresholds to sign, and the main challenge to a petition organizer is reaching these individuals in the crucial time window to be successful. 

Although we do not 
find evidence of the S-shaped curve identified by Gonzalez-Bailon et al.~\cite{gonzalez2011}, our pattern of early growth could be commensurate with the finding 
of others that social contagion (where social contagion refers to dissemination) requires exposure to a diversity of sources \cite{centola2007}, and that evidence of 
recruitment bursts would suggest that the effect of multiple and diverse exposures are magnified if they take place in a short time window \cite{gonzalez2011}. These observations 
suggest that S-shaped commutative growth might not be a proper model to describe Internet-based collective behavior 
where the costs of participation are very low and the pool of potential participants is very large. 
Instead, an initial momentum that decays over time seems to be a more relevant picture for online mobilization.
Further research would be needed 
to investigate this hypothesis, examining the spread of petitions through social networking sites to understand variations in the dynamics of different platforms.


\begin{backmatter}

\section*{Competing interests}
The authors declare that they have no competing interests.

\section*{Author's contributions}
TY has performed the data analysis and modeling. SH has collected the data. HM designed the research project. All the authors have contributed to writing and reviewed the manuscript.

\section*{Acknowledgements}
  This work was supported by the Economic and Social Research Council [grant number RES-051-27-0331].
 

\bibliographystyle{bmc-mathphys} 
\bibliography{references}       

\end{backmatter}
\end{document}